\documentclass[10pt,conference,letterpaper]{IEEEtran}
\usepackage{times,amsmath,epsfig}
\IEEEoverridecommandlockouts

\title{INSQ: An Influential Neighbor Set Based Moving \emph{k}NN Query Processing System}

\author{\IEEEauthorblockN{Chuanwen Li\IEEEauthorrefmark{1},
Yu Gu\IEEEauthorrefmark{1},
Jianzhong Qi\IEEEauthorrefmark{2},  
Ge Yu\IEEEauthorrefmark{1},  
Rui Zhang\IEEEauthorrefmark{2} and
Qingxu Deng\IEEEauthorrefmark{1}}
\IEEEauthorblockA{\IEEEauthorrefmark{1}College of Information Science and Engineering, 
Northeastern University\\Email: \{lichuanwen,guyu,yuge,dengqx\}@ise.neu.edu.cn 
\IEEEauthorblockA{\IEEEauthorrefmark{2}Department of Computing and Information Systems, 
University of Melbourne
\\ Email: \{jianzhong.qi,rui.zhang\}@unimelb.edu.au}}}

\usepackage{graphicx}
\usepackage{balance}
\usepackage{xspace}
\usepackage{xcolor}
\usepackage{mathptmx}
\usepackage{amsmath, amssymb}
\usepackage{epsfig}
\usepackage{subfigure} 
\usepackage{url}
\usepackage[breaklinks]{hyperref}
\usepackage[T1]{fontenc}
\usepackage{color}
\usepackage[mathcal]{euscript}
\usepackage{array}

\usepackage[lined, ruled,vlined]{algorithm2e}
\DontPrintSemicolon 
\LinesNumbered

\usepackage{paralist} \let\itemize\compactitem
\let\enditemize\endcompactitem 
\let\endenumerate\endcompactenum

\makeatletter
\newcommand{\figcaption}{\def\@captype{figure}\caption}
\newcommand{\tabcaption}{\def\@captype{table}\caption}
\makeatother

\newtheorem{definition}{Definition}
\newtheorem{theorem}{Theorem}

\newtheorem{lemma}{Lemma}

\newcommand{\eat}[1]{}

\newcommand{\knn}{\emph{k}NN\xspace}
\newcommand{\mknn}{M\emph{k}NN\xspace}

\newcommand{\vn}[2]{\ensuremath{N_{#2}(#1)}}
\newcommand{\as}[1]{\ensuremath{NN_k(#1)}}

\newcommand{\vorn}[2]{\ensuremath{V^{#2}(#1)}}
\newcommand{\vornt}[3]{\ensuremath{V^{#3}_{#2}(#1)}}

\newcommand{\dn}[1]{{\ensuremath{d}{(#1)}}\xspace}
\newcommand{\lb}{\linebreak[0]}

\newcommand{\is}{\ensuremath{\circlearrowleft}}
\newcommand{\bs}[1]{\ensuremath{b(#1)}}
\newcommand{\okv}{order-$k$ Voronoi\xspace}
\newcommand{\mis}[1]{\ensuremath{MIS(#1)}}

\newcommand{\rk}{\ensuremath{\lfloor \rho k \rfloor}\xspace}

\begin{document}

\IEEEpubid{\makebox[\columnwidth]{\hfill Copyright~\copyright~2016 IEEE}
\hspace{\columnsep}\makebox[\columnwidth]{}
}

\maketitle

\begin{abstract}


We revisit the moving $k$ nearest neighbor (\mknn) query, which computes one's $k$ nearest neighbor set and maintains it while at move. Existing \mknn algorithms are mostly safe region based, which lack efficiency due to either computing small safe regions with a high recomputation frequency or computing larger safe regions but with a high cost for each computation.
In this demonstration, we showcase a system named INSQ that adopts a novel algorithm called the Influential Neighbor Set (INS) algorithm to process the \mknn query in both two-dimensional Euclidean space and road networks.  This algorithm uses a small set of safe guarding objects instead of safe regions.  As long as the the current $k$ nearest neighbors are closer to the query object than the safe guarding objects are, the current $k$ nearest neighbors stay valid and no recomputation is required. Meanwhile, the region defined by the safe guarding objects is the largest possible safe region. This means that the recomputation frequency is also minimized and hence, the INS algorithm achieves high overall query processing efficiency.


\end{abstract}

\section{Introduction}

Location-based services (LBS) such as navigation services~\cite{DBLP:conf/icde/AljubayrinQJZHW15} are more and more popular as smart mobile devices have become prevalent. On-going efforts have been made to improve the user experience of mobile LBS through improvements in moving query processing efficiency. In this paper we revisit a major type of moving query, the \emph{moving $k$ nearest neighbor (\mknn)} query~\cite{z1, z2}. \emph{Given a moving query object and a set of static data objects, the \mknn query reports the $k$ nearest neighbors of the query object continuously while it is moving}. For example, a \mknn query can be used to report the 3 nearest gas stations continuously while one drives on a highway, or the 5 nearest points of interest (POI) continuously while a tourist is walking around a city.

Studies on the \mknn query have mainly used the \emph{safe region based} approach~\cite{DBLP:journals/kais/Li0QZY15,VSTAR, okabe2009spatial}. The idea is based on that objects at nearby positions often share the same $k$NN set. There may be a region where all points have the same \knn set. Inside such a region, an object is ``safe'' to move freely without causing its \knn set to change. Thus, such a region is called a safe region. Using the safe region significantly reduces the \mknn query processing cost because the \knn set will only need to be recomputed when the query object moves out the safe region.  However, the safe region also brings in some overhead: (i) \emph{construction overhead}: the safe region needs to be recomputed every time the \knn set is recomputed; (ii) \emph{validation overhead}: whether the query object is still inside the current safe region needs to be checked every time the query object location updates.


Existing methods either fall short in construction overhead or validation overhead. Specifically, earlier studies~\cite{GIScience06,okabe2009spatial} use \emph{Voronoi cells}~\cite{okabe2009spatial} as the safe regions, which have high construction overhead.  A more recent study~\cite{VSTAR} uses relaxed safe regions to reduce the construction overhead, but it has to recompute the safe regions more frequently and has higher validation overhead.

A Voronoi cell based safe region is essentially an \emph{order-$k$ Voronoi cell}. Given a \knn set, its order-$k$ Voronoi cell is defined as a region where the \knn set stays valid~\cite{okabe2009spatial}. The computation cost of computing order-$k$ Voronoi cells on the fly is prohibitively high since it requires a combination of $k$ order-1 Voronoi cells. Precomputing the order-$k$ Voronoi cells~\cite{GIScience06} is also unpractical due to the rapid increase in the number of order-$k$ Voronoi cells as $k$ increases.



In this demonstration, we showcase an Influential Neighbor Set based system named \emph{INSQ}. The influential neighbor set (INS) is an algorithm to avoid the high cost of \mknn query processing based on our previous paper~\cite{chiewenVLDB}, which uses \emph{safe guarding objects} instead of safe regions. The intuition is that, since the query object moves continuously, when the $k$NN set changes, some data object near the current $k$NN objects must become one of the new $k$NNs. We call this type of data objects (i.e., data objects near the current $k$NN objects) the safe guarding objects. As long as no safe guarding object becomes a $k$NN, the current \knn set stays valid and does not need recomputation. Conceptually, the safe guarding objects define a safe region as large as the order-$k$ Voronoi cell. Thus, they guarantee minimum $k$NN set recomputation and hence minimum communication between the query client and the query processor, which is critical in LBS. To improve efficiency we identify a special set of safe guarding objects and call it the \emph{influential neighbor set (INS)}, which can be computed and validated in time linear to $k$. This set reduces both construction and validation overhead at the same time. 

We build an interactive demonstration program that shows the internal mechanism of the INS algorithm. It simulates a moving query object and displays the changing $k$NNs as well as the safe guarding objects continuously. The users are given options to customize several parameters for information and aesthetic purposes such as the number of nearest neighbors displayed. \emph{Note that in addition to the two-dimensional Euclidean space scenario which was discussed in our previous paper~\cite{chiewenVLDB}, we also discuss and 
demonstrate how the INS algorithm works in road networks in this paper.}

\section{Influential Neighbor Set}
\label{sec:l}

The key idea of our \mknn query processing approach is to use a set of \emph{safe guarding objects}. As long as the query object is closer to the current $k$NNs than the safe guarding objects, it is guaranteed that the current \knn set is still valid, and therefore, we do not need to recompute the \knn set.

We call the set of safe guarding objects an \emph{influential set (IS)}, in the sense that they are \emph{influential} in determining whether a \knn set is valid. The \emph{influential neighbor set (INS)} is a special IS that can be efficiently computed, which will be detailed below. Now we first formalize the IS.
\begin{definition} (Influential Set, IS)
  Given a set of data objects $O$, a query object $q$ and a \knn set $O' \subset O$, we call a set $S \subseteq O\backslash O'$ an \emph{influential set} of $O'$, denoted by $S \is O'$, if $O'$ stays as the \knn set of $q$ as long as the objects in $O'$ are closer to $q$ than the objects in $S$ are, i.e.,
  \begin{equation}
    \label{eq:19}
     O' = \as{q} \iff O' \prec_{q} S.
  \end{equation}
\end{definition}

\vspace{-5pt}
Here,  $\as{q}$ is a function that returns the \knn set of $q$, and $A \prec_q B$ denotes that any object in a set $A$ is closer to $q$ than every object in a set $B$.

\begin{figure}[!hbp]
  \centering
  \includegraphics[width=0.2\textheight]{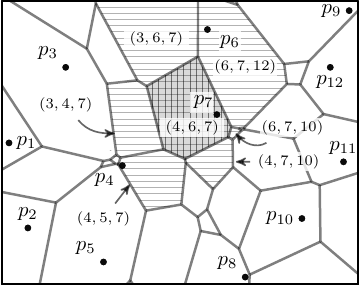}
  \vspace{-2pt}
  \caption{The minimal influential set (MIS) of $O'=\{p_4,p_6,p_7\}$} 
\label{fig:mis}
\vspace{-3pt} 
\end{figure}

Trivially, $O\backslash O'$ is an IS of $O'$, since by definition the \knn set $O'$ is closer to $q$ than $O\backslash O'$. However, using this set to validate the \knn set is expensive.
We aim to find the \emph{minimal influential set}, which is the smallest influential set that can guarantee the validity of the current \knn set. 

\begin{definition}(Minimal Influential Set, MIS)\label{df:mis}
  Given a \knn set $O'$, the \emph{minimal influential set} (MIS) of $O'$ is
  \begin{equation}
    \label{eq:1}
    \mis{O'}= (\bigcup_{\vorn{O'}{k} \cap \vorn{O''}{k} \ne \emptyset} O'') \backslash O'.
  \end{equation}
\end{definition}

\vspace{-5pt}
Here, $\vorn{O''}{k} \cap \vorn{O'}{k} \ne \emptyset$ means that the two \okv cells are neighboring cells, i.e., they share an edge. Figure~\ref{fig:mis} gives an example, where $O'=\{p_4,p_6,p_7\}$ and the cross-lined region denotes $\vornt{O'}{O}{3}$. There are five neighboring order-3 Voronoi cells, as denoted by the horizontal-lined regions. The triples associated with the neighboring cells denote the corresponding objects, e.g., $(3,4,7)$ is associated to $V^3_O(p_3,p_4,p_7)$. The union of the triples excluding $O'$ is $3, 5, 10, 12$.  Therefore, the MIS of $O'$, $\mis{O'}=\{p_3, p_5, p_{10}, p_{12}\}$. 

In~\cite{chiewenVLDB} we prove that MIS is minimal. However, similar to computing the strict safe region (i.e., an order-$k$ Voronoi cell), computing the MIS is an expensive operation. To reduce the computational cost, we compute an influential set that is slightly larger than the MIS but can be computed much more efficiently. By definition of the influential set, this will not affect the strictness in validating the \knn sets. 

The influence set we use is the \emph{influential neighbor set (INS)}, which is defined based on the Voronoi neighbors.

\begin{definition} (Voronoi neighbor set)
  Given a Voronoi diagram on data object set $O$, we call an object $p_j$ a \emph{Voronoi neighbor} of another object $p_i$ if the two Voronoi cells of the two objects share an edge, i.e., $V(p_i) \cap V(p_j) \ne \emptyset$. We call $O' \subset O$ that contains all Voronoi neighbors of $p_i$ the \emph{Voronoi neighbor set} of $p_i$, denoted by $\vn{p_i}{O}$.  
\end{definition}

For an order-1 Voronoi diagram, the Voronoi neighbor sets can be precomputed and stored with little overhead~\cite{sharifzadeh2010vor}. \emph{The influential neighbor set is the union of the order-1 Voronoi neighbor sets of all current $k$NNs.}

\begin{definition}\label{df:in} (Influential neighbor set, INS)
  Given a \knn set $O'$, an object $p$ is an \emph{influential neighbor} of $O'$ if $p$ is not in $O'$ while it is a Voronoi neighbor of an object $p'$ in $O'$, i.e., for $p \notin O', \exists p' \in O': V(p) \cap V(p') \ne \emptyset$.  We call the set of all influential neighbors of $O'$ the \emph{influential neighbor set (INS)} of $O'$, denoted by $I(O')$,
\begin{equation}
    \label{eq:5}
    I(O') =  (\bigcup_{p' \in O'} \vn{p'}{O}) \backslash O'.
  \end{equation}
\end{definition}
\vspace{-5pt}

In~\cite{chiewenVLDB} we prove that an INS is an IS, which guarantees the correctness of using the INS for query processing.

\section{Query Processing} \label{sec:m}

 When a \mknn query is issued, we compute an initial \rk NN set, denoted by $R$, and the INS of $R$, $I(R)$. Here, $\rho \geq 1$ is a system parameter to balance the query result communication and recomputation costs. We call it the \emph{prefetch ratio} and have discussed how to choose its value in~\cite{chiewenVLDB}. To improve the efficiency of computing $I(R)$, we precompute the Voronoi diagram of $O$ and index it with an VoR-tree~\cite{sharifzadeh2010vor}. 
 
We return the set $R$ and $I(R)$, where the top $k$ objects of $R$ are marked as the \knn set (\as{q}) and $I(R) \cup R \backslash \as{q}$ are marked as the IS.  Then query maintenance starts. At each timestamp, we check whether the current \knn set is nearer to $q$ than the IS: (i) if it is, then the current \knn set is still valid; (ii) if it is not, we perform the \knn set update procedure. If there are data object updates, we also update the \knn set and the IS according to the data object updates. 

\eat{
\begin{algorithm}[!htbp]
\caption{Query Maintenance}\label{alg:srv}

\SetKwInOut{Input}{input}
\SetKwInOut{Output}{output}
\SetKwFunction{KwNear}{Validation}
\SetKwFunction{KwSwitch}{Update}
\SetKwFunction{KwInsertion}{Insertion}
\SetKwFunction{KwDeletion}{Deletion}

\Input{query object $q$, prefetched set $R$, current \knn set $O'$ and its influential set $IS$}
\Output{\knn set at each timestamp}

\While{true}{
  $r \leftarrow $\KwNear{$q$, $O'$, $IS$}\;
  \If{$r.isValid = $ false}{
    \KwSwitch{$q$, $R$, $O'$, $IS$, $r.candidate$, $r.delete$}\;
  }
}
\end{algorithm}
}
\subsection{kNN Set Validation}
\label{sec:order-checking}

To validate the current \knn set, we scan the \knn set and the IS. In the \knn set we find the one that is the farthest from $q$, denoted by $r.delete$; in the IS we find the one that is the nearest to $q$, denoted by $r.candidate$. If $r.candidate$ is closer to $q$ than $r.delete$, the current \knn set has become invalid and we need to update it.  

\subsection{kNN Set Update}
\label{sec:result-set-switching}

There are two cases for \knn set update: (i) If $q$ has entered a neighboring order-$k$ Voronoi cell, the new \knn set will differ from the current one by only one data object. In this case, we do not need to recompute the entire \knn set but can use the existing \knn set to compose the new \knn set. (ii) If $q$ is not in a neighboring order-$k$ Voronoi cell, we first check whether the new \knn set is still in $R$. If yes, then we just need to return this new \knn set. If not, we compute the new sets of $R$ and $I(R)$ and return the new \knn set and IS.

\section{INS in Road Networks}
\label{sec:ins-road-networks}

We extend the influential neighbor set to road networks in this demonstration paper. We consider a planar undirected (connected) graph $G = \langle V, E \rangle$ formed by a set of vertices $V=\{v_1, v_2, \ldots, v_m\}$ and a set of edges $E=\{e_1,\ldots,e_l\}$. We assume that the data objects $O=\{p_1,\lb \ldots,p_n\}$ are all at the vertices (otherwise we can add them to the set of vertices and add edges accordingly).

\begin{figure}[!ht]
  \centering  
  \subfigure{\includegraphics[width=0.22\textheight]{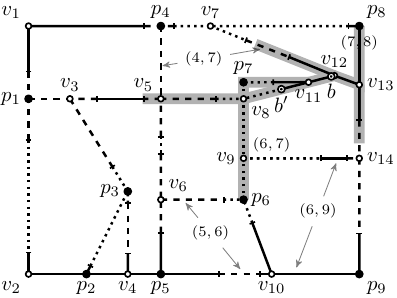}}
  \caption{A road network example}
  \label{fig:t9}
\end{figure}

The concept of Voronoi diagram still applies in road networks. 
As such, we can show that the influential neighbor set defined by Definition~\ref{df:in} is still a super set of the MIS defined by Definition~\ref{df:mis} in road networks. 

\begin{theorem}\label{thm:ri}
  In a road network, given a \knn set $O_{knn}$, $I(O_{knn})$ as defined by Definition~\ref{df:in}, and $MIS(O_{knn})$ as defined by Definition~\ref{df:mis}:
  \begin{equation}
    \label{eq:21}
    MIS(O_{knn}) \subseteq I(O_{knn}).
  \end{equation}
\end{theorem}
\vspace{-5pt}

The detailed proof is omitted due to space limit. We use Fig.~\ref{fig:t9} to illustrate the idea. The figure shows an order-2 Voronoi diagram in a road network. We denote the vertices with data objects by solid dots and the rest of the vertices by hollow dots, respectively.  We denote the edge segments that belong to the same order-2 Voronoi cell by the same line type. There is a number pair, e.g., $(6,7)$, besides these edge segments, meaning that these edge segments belong to an order-2 Voronoi cell, e.g., $V^2(p_6, p_7)$.

Let the current $k$NN set be $O_{knn} = \{p_6, p_7\}$. The MIS by definition should be $MIS(O_{knn})=\{p_4, p_5, p_8, p_9\}$. 
For each pair of objects $(p, p') \in O_{knn} \times MIS(O_{knn})$, e.g., ($p_7, p_8$), we can find a ``\emph{mid-point}'' on the shortest path between $p_7$ and $p_8$.
For example, the mid-point between $p_7$ and $p_8$ is denoted by $b$ in the figure. This mid-point satisfies $\dn{b, p_7} = \dn{b, p_8}$; no other object in $O\backslash O_{knn}\backslash I(O_{knn})$ is nearer to $b$ than $p_7$ or $p_8$. Here, $\dn{}$ denotes the shortest road network distance between two objects. The existence of this mid-point indicates that $p_7$ and $p_8$ are also order-1 Voronoi neighbors, which means that $p_8$ should also belong to $I(O_{knn})$. Thus, we have shown that every object in $MIS(O_{knn})$ also belongs to $I(O_{knn})$, i.e., $MIS(O_{knn}) \subseteq I(O_{knn})$.

The above theorem allows to validate the $k$NN set using the INS in road networks. However, in road networks, checking whether the query object is nearer to any object in the INS than the objects in the \knn set is not a trivial operation. This is because it requires network distance (shortest path) computation. 
To constraint the search space of this computation, we have proven the following theorem.

\begin{theorem}\label{thm:noshorter}
  Given a Voronoi diagram $D_O$ on a road network with a set of data objects $O$, a set $O_{knn} \subset O$ and its INS $I(O_{knn})$, and a Voronoi diagram $D_{O_{knn}\cup I(O_{knn})}$ formed by the edges and vertices from the Voronoi cells of the objects in $O_{knn}$ and $I(O_{knn})$, if the \knn set of a query object $q$ is $O_{knn}$ on $D_{O_{knn}\cup I(O_{knn})}$, then the \knn set of $q$ on $D_O$ is also $O_{knn}$.
\end{theorem}

This theorem guarantees that we just need to consider the (smaller) road network formed by the current $k$NN set and the INS for query validation. We omit the detailed proof 
due to space limit. 
Based on the two theorems above, we have developed the INS algorithm for road networks which is showcased in the following section.

\section{Demonstration}
\label{sec:prog}

The system\footnote{Source code available at \url{https://github.com/chiewen/CkNN.git}} is implemented as a Scala Swing application. 
The application runs in two modes, \emph{Road Network} mode and \emph{2D Plane} (Euclidean space) mode. 
Its user interface consists of two panels (cf. Figure~\ref{fig:demo_whole}):

(i) \emph{Control panel.} This is the left panel of the user interface. It is used for setting the demonstration parameters, which include the \emph{Global Setting}, the \emph{Road Network} setting and the \emph{2D Plane} setting. 

The global setting includes the underlying map to be used, the data space to be used, and the value of the query parameter $k$. There are also two buttons ``Save'' and ``Read'' which are for recording and loading the  demonstration settings. 

The road network setting controls the operations being performed: ``\emph{Node}'', ``\emph{Site}'', ``\emph{Trajectory}'', and ``\emph{Demo}''. When one of the first three options is chosen, 
users can add/move/delete the network nodes/data objects/query trajectory. When ``Demo'' is chosen, a M$k$NN query will be simulated. The setting also includes options to control which objects to be shown as well as the query object's moving speed in the simulation.

The 2D plane setting includes the number of data objects to generate ($n$), the value of the prefetch ratio ($\rho$), and whether to show the Voronoi cells and the corresponding safe regions.

Under these settings, the current \knn set and the INS are displayed.

\def\wdh{.36}
\begin{figure}[!hbpt]
  \centering
    \includegraphics[width=\wdh\textheight]{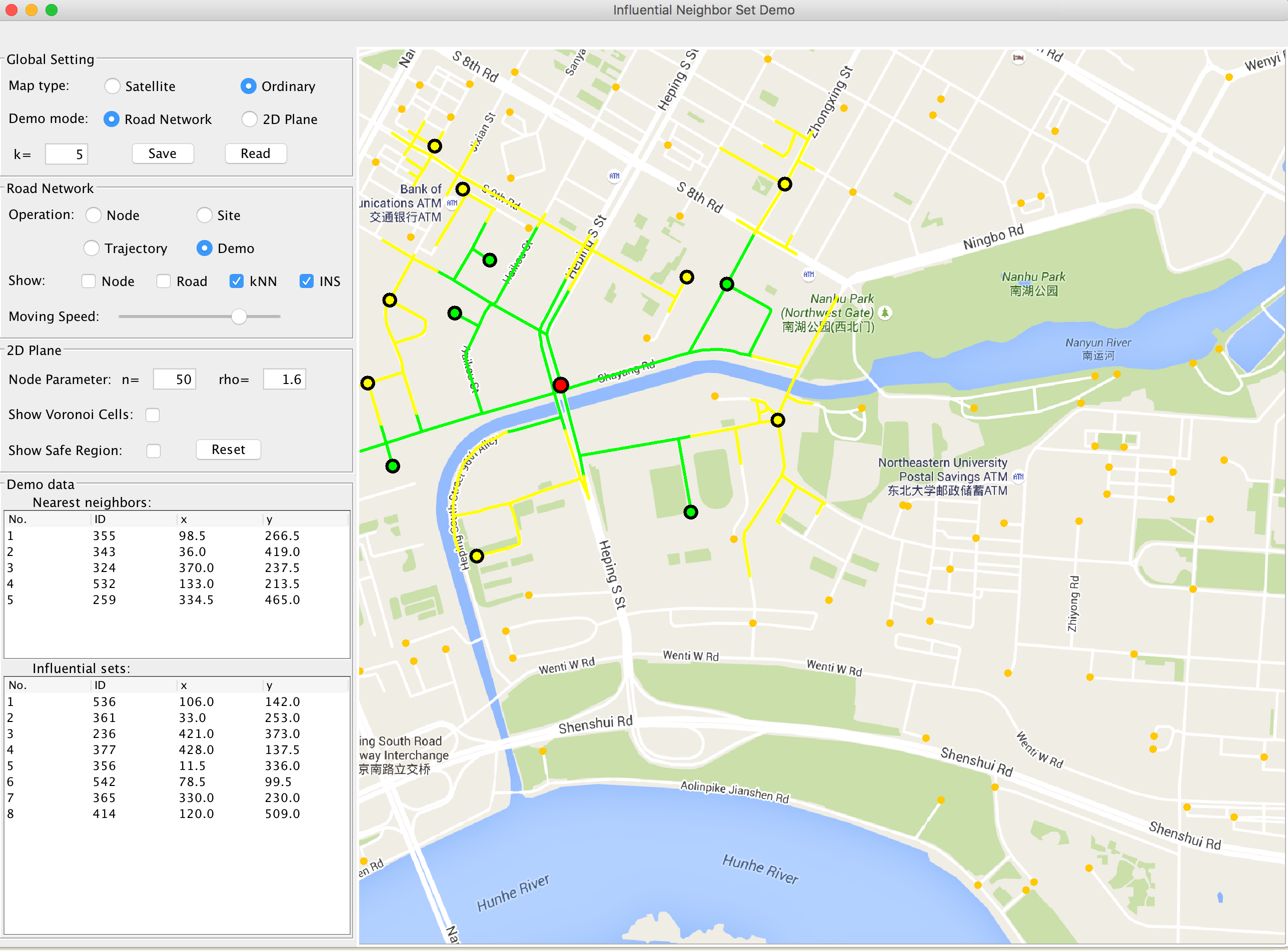}
    \vspace{-2pt}
    \caption{Screenshots (Road Network, $k=5$)}
    \vspace{-10pt}
\label{fig:demo_whole} 
\end{figure}

(ii) \emph{Main panel.} This is the right panel of the user interface. It displays a map of a road network where data objects (orange dots) and the query object (red dot) can be placed onto. Users can specify a trajectory for the query object. In the Road Network mode, this trajectory must confine the underlying road network; in the 2D Plane mode, the trajectory can have any shape. When query simulation starts, the query object will move along the specified trajectory. The INS algorithm will run to compute the $k$NN set and the INS set, which are shown as the green dots and the yellow dots, respectively.

In the Road Network mode, the Voronoi cells of the objects in the \knn set and the INS are represented by the sets of  green and yellow network edges, respectively.  

In the 2D Plane mode (Fig.~\ref{fig:demo}), the data objects are surrounded by their respective order-1 Voronoi cells. The ones enclosed by green squares represent the objects in $R$. We use the cyan polygon to represent the current order-\emph{k} Voronoi cell, which will turn red when it becomes invalid. We circle two special objects, the farthest object to $q$ in the \knn set and the nearest object to $q$ in the INS, with a green circle and a red circle passing through them, respectively, where the center of both circles are at $q$.  These two objects are special in the sense that the green circle should always enclose all the objects in the \knn set, while the red circle should never enclose any object in the INS, as long as the current \knn set  is still valid.

\def\wdh{.25}
\begin{figure}[!hbpt]
  \centering
    \subfigure[\scriptsize The \knn set is valid]{\includegraphics[width=\wdh\textheight]{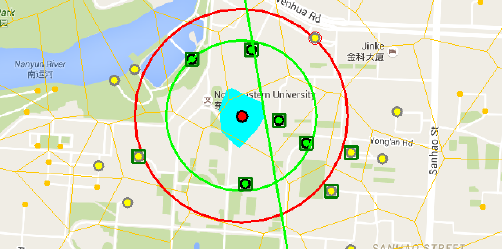}}
    \subfigure[\scriptsize The \knn set is invalid]{\includegraphics[width=\wdh\textheight]{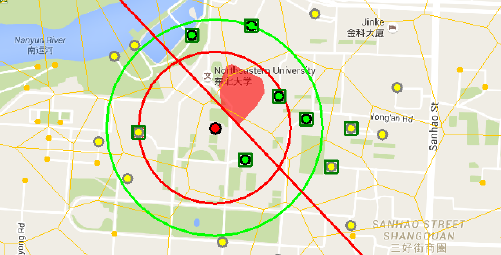}}
    \vspace{-2pt}
    \caption{Screenshots (2D plane, $k=5$ and $\rho=1.6$)} 
    \vspace{-10pt}
\label{fig:demo} 
\end{figure}

\textbf{Demonstrated Scenarios:} We take the 2D Plane mode as an example. Fig.~\ref{fig:demo} displays two screenshots of the demonstration program, where each shows: (i) the query object stays in the order-\emph{k} Voronoi cell of the current \knn set (i.e., the \knn set is valid), and (ii) the query object has moved out of the order-\emph{k} Voronoi cell (i.e., the \knn set is invalid). 

When query $q$ moves out of the order-\emph{k} Voronoi cell in Fig.~\ref{fig:demo} (a) and moves into the position shown in Fig.~\ref{fig:demo}~(b), the current \knn set becomes invalid since the green circle is now enclosed by the red circle (i.e., the farthest object to $q$ in the \knn set is farther to $q$ than the nearest object to $q$ in the INS). 

\vspace{-5pt}
\section{Conclusion}
\label{sec:conclusion}
We built and showcased a system named INSQ to process the M$k$NN query on both Euclidean space and road networks, based on a highly efficient algorithm INS. The program simulates the movement of the query object along the  trajectory and displays the changing $k$NNs and the corresponding influential neighbors which are the key feature of the INS algorithm. Users can customize the behavior of the simulation and observe the mechanisms of the algorithm in detail. The source code of the INS algorithm and the INSQ system is publicly accessible through the Internet, which can be easily adapted to real systems such as smart phone applications.

\section*{Acknowledgements}
\label{sec:acknowledgement}
This work is supported by the National Natural Science Foundation of China under Grant No.\ 61300021, 61472071, 61472072, 61402155, the National Basic Research Program of China (973 Program) under Grant No.\ 2012CB316201 and 2014CB360509, the Fundamental Research Funds for the Central Universities of China No.\ N140404008.
Jianzhong Qi is the recipient of a Melbourne School of Engineering Early Career Researcher Grant (project reference number 4180-E55), and a University of Melbourne Early Career Researcher Grant (project number 603049). Rui Zhang is supported by ARC Future Fellow project FT120100832.

\begin{small}
\bibliographystyle{abbrv}
\bibliography{knn} 
\end{small}

\balance
\balance

\end{document}